\documentclass[runningheads]{llncs}
\usepackage{graphicx}
\usepackage{url}
\begin{document}
\title{Cyber-Physical Platform for Preeclampsia Detection}
%
%


\author{Iuliana Marin\inst{1}\orcidID{0000-0002-7508-1429} \and
Maria Iuliana Bocicor\inst{1}\orcidID{0000-0003-2374-9551} \and
Arthur-Jozsef Molnar\inst{1}\orcidID{0000-0002-4113-2953}}
\authorrunning{Marin et al.}
%
\institute{SC Info World SRL, Bucharest, Romania
\email{\{iuliana.marin,iuliana.bocicor,arthur.molnar\}@infoworld.ro}\\
\url{http://www.infoworld.ro/home/}}

\maketitle              
\begin{abstract}
Hypertension-related conditions are the most prevalent complications of pregnancy worldwide. They manifest in up to 8\% of cases and if left untreated, can lead to serious detrimental effects. Early detection of their sudden onset can help physicians alleviate the condition and improve outcomes for both would-be mother and baby. Today's prevalence of smartphones and cost-effective wearable technology provide new opportunities for individualized medicine. Existing devices promote heart health, they monitor and encourage physical activity and measure sleep quality. This builds interest and encourages users to require more advanced features. We believe these aspects form suitable conditions to create and market specialized wearable devices. The present paper details a cyber-physical system built around an intelligent bracelet for monitoring hypertension-related conditions tailored to pregnant women. The bracelet uses a microfluidic layer that is compressed by the blood pressing against the arterial wall. Integrated sensors register the waveform and send it to the user's smartphone, where the systolic and diastolic values are determined. The system is currently developed under European Union research funding, and includes a software server where data is stored and further processing is carried out through machine learning.

\end{abstract}

\section{Introduction}
\label{sec:introduction}
Hypertensive disorders affect up to 8\% of all pregnancies \cite{kongwattanakul2018incidence}, with increased prevalence in women who already carried a preeclamptic pregnancy. In the United States, hypertension related conditions affect between 7 to 15\% of pregnancies \cite{2}. They are classified into chronic hypertension, gestational hypertension and preeclampsia \cite{12,15}, and represent the most common medical problem encountered during pregnancy. Chronic hypertension is present before 20 weeks of pregnancy, or in cases where the woman is already on anti-hypertensive medication. Gestational hypertension is diagnosed after 20 weeks of pregnancy. When combined with proteinuria (the presence of protein in urine), or organ dysfunction such as renal or liver involvement, it is diagnosed as preeclampsia. If left untreated, preeclampsia can lead to adverse effects for both the mother and baby, including restricted fetal growth, organ damage, and seizures (eclampsia), which necessitate delivering the baby to prevent further consequences \cite{12}. 

Early detection allows medical intervention with the aim of maintaining blood pressure under a safe threshold, managing the time of delivery for minimal risk to both mother and baby as well as prophylactic use of magnesium sulphate during labor \cite{4,5}.

In the last decade, medicine is going through an important transformation fueled by the ubiquity of wearables, together with increased awareness and interest from the general public \cite{1}. We believe this can be harnessed to improve the outcomes for conditions such as preeclampsia, which are well known, but under reported in many parts of the globe. In one region, 77\% of pregnant women affected were unaware of the condition or its possible consequences \cite{13}. 

In this paper we detail a cyber-physical system that targets early detection of preeclampsia and other hypertension conditions related with pregnancy \cite{16}. The system is being developed under funding from the European Union. It integrates a custom developed wearable bracelet for real-time measurement of blood pressure with a smartphone application and a software server. 

The system connects the would-be mother with her clinician and provides an integrated platform to alert clinicians at the onset of persistent hypertension \cite{17}. As preeclampsia can have sudden onset and is difficult to manage, registering its early warning signs allows more flexibility in its diagnosis and management \cite{4}.

One of the principal innovations regards the sensor-driven bracelet used to monitor blood pressure. Current products are based on infrared or oximeter sensors that directly measure pulse and blood oxygen levels, which they then correlate with blood pressure \cite{9,carek2017seismowatch,11}. They enable taking measurements at shorter intervals and without so much preparation \cite{carek2017seismowatch}. However, these devices can require calibration and they use approximation to estimate blood pressure. The proposed solution increases accuracy by measuring blood pressure directly and eliminates calibration requirements.

The wrist-worn bracelet is built around a diminutive sensor comprised of a microfluidic layer located between two sensing membranes. External pressure of the blood pushing on the arterial wall is picked up by the microfluidic layer and results in a change of impedance which leads to the complete signal of the waveform being recorded. Readings are transmitted to the wearer's smartphone using an integrated wireless module with a BLE chip. The energy efficiency of BLE and the sensor allow the wrist bracelet to measure blood pressure several times during an hour and transmit data without requiring frequent charging. Reading history and trends can be consulted using the smartphone application. They are also forwarded to the software server, which clinicians and end-users can access over the web. When hypertension is detected, the associated clinician receives an application alert; this helps with continuously monitoring patients and lower the time to intervention.

\section{Related Work}
\label{sec:relatedwork}
Hypertension is a major cause of premature death worldwide \cite{who2017} and an important cause of complications during pregnancy \cite{braunthal2019hypertension}. The idea of continuous measurement of blood pressure outside of medical facilities has only recently been adopted and during the past years various applications and devices have been developed for this purpose. One such device is H2-BP, a wearable blood pressure monitor in the form of a light wrist band \cite{h2care}. It measures blood pressure with an accuracy of 5mmHg every 30 - 50 seconds, as well as heart rate with an accuracy of $\pm5\%$. While measuring, the indication is to hold a correct position, in which the device is at heart level, and not to move or speak \cite{h2care}. An Android application is also available and communication between the device and smartphone is achieved via Bluetooth.  


Omron’s HeartGuide Blood Pressure Watch is an oscillometric wrist wearable blood pressure monitor composed of miniaturised components for traditional oscillometric measurement \cite{omron}. It includes an inflating cuff and can take a blood pressure reading in approximately 30 seconds. Reading is triggered by pressing a button and lifting the arm to heart level, where it must be kept still during the measurement. The company also offers the Heart Advisor smartphone application, which synchronises readings to a cloud service and optionally with Apple HealthKit. It provides features for monitoring blood pressure, activity and sleep quality and it allows for configuration of reminders and settings.

Unlike the devices presented above, the SeismoWatch introduced by Carek et al. \cite{carek2017seismowatch} measures blood pressure via a novel technique which employs the theory of pulse transit time. The device they propose is a wrist watch that must be held against the sternum for less than 15 seconds to detect micro-vibrations of the chest wall associated with the heartbeat. Blood pressure is estimated by considering its inverse correlation with the time interval for a pulse wave to travel from a point close to the heart to a more distant location along the arterial tree (e.g. the wrist). An accelerometer and an optical sensor on the watch measure the pulse wave travel time and according to this the device estimates systolic and diastolic blood pressure. 
A similar idea is used by Hsu et al. \cite{hsu2014skin}, who propose a non-invasive method of measuring blood pressure: two pressure sensors simultaneously detecting blood pressure waveforms are placed on the person's neck and wrist and the delay time between them is used to determine the local pulse waveform. The skin-coupled wearable system can continuously capture blood pressure waveforms and the experiments showed that the prototype was capable of detecting high-fidelity blood pressure waveforms.


A solution specifically tailored for preeclampsia uses an F1 smart wristwatch which measures blood pressure and sends the triggered values via Bluetooth to a mobile application installed on the user's cell phone \cite{musyoka2019}. The bracelet model was selected after comparison with other wristwatch models where it was found that taking measurements from users with dark complexion was difficult, resulting in finger tip measurements to determine blood pressure \cite{musyoka2020}. The mobile application is intended to be helpful for the expectant mother and her caregiver. This application includes a maintained database comprising gathered records. The caregiver is alerted if the expectant mother is in a critical circumstance.

A device utilized for dealing with the health and well-being during pregnancy is the Ava bracelet \cite{ava}. It determines cycle, fertility, sleep quality and pregnancy based on physiological signals which are collected automatically while asleep, at night time. At the point when the user awakens, the bracelet is synchronized to the mobile application and the algorithm of Ava shows the outcomes. Its mobile application displays the information about pulse and breathing rates, skin temperature, and heart rate variability ratio which determines physiological stress. Sleep is checked by the Ava bracelet and it calculates complete rest time, percentage of light contrasted with deep and REM sleep, including the rest patterns. At the point when the woman is pregnant, the mobile application can screen her weight and show the development on its graphical user interface. The mobile application monitors and informs its user about the baby's advancement during pregnancy.

When compared with existing solutions, our proposed wearable-based system has clear advantages: first, it automatically measures blood pressure without the need to maintain a certain position, it does not use inflatable cuffs, and it is comfortable and easy to wear continuously. Measurement accuracy is ensured by the design of the device, in which the sensor covers a large surface of the inner wrist. Furthermore, the software server employs machine learning techniques to detect potential signs of preeclampsia \cite{17}. Considering the above, the system we propose is original, technologically advanced and accessible both with regard to cost and ease of use.

\section{Technological Solution for Preeclampsia Prevention}
\label{sec:introduction}
\begin{figure}[h]
    \centering
    \includegraphics[width=\linewidth]{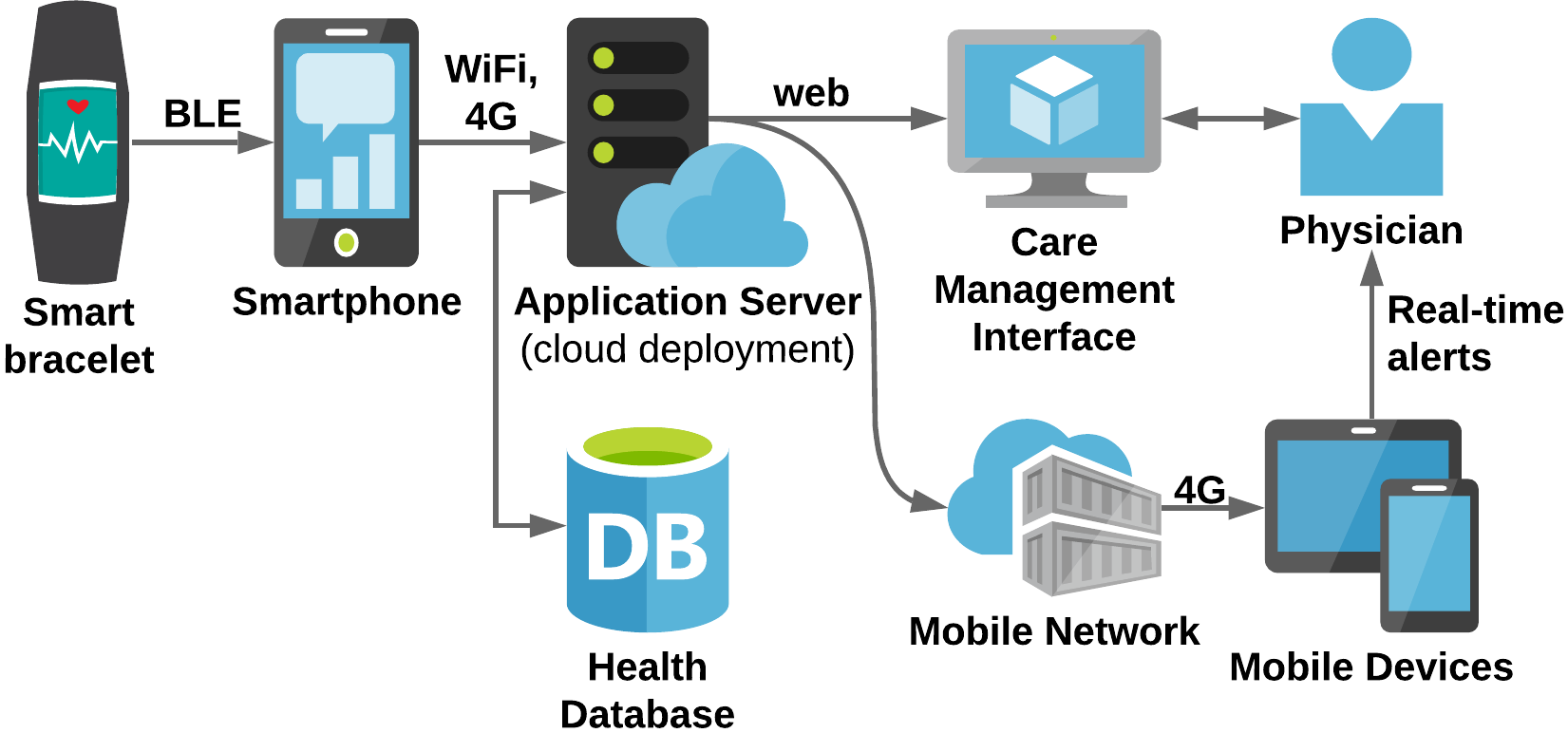}
    \caption{High-level architecture of the platform}
    \label{fig:architecture}
\end{figure}

The platform provides an end-to-end solution for monitoring the onset and progression of hypertension-related conditions tailored for pregnant women. In addition to the custom-developed smart bracelet, it incorporates one of the end-user's smart mobile devices. A custom developed application is installed on this device, which is usually an Android or iOS powered smartphone or tablet. It is used to communicate with the bracelet using the low power BLE standard, process raw readings and transmit them to the software server. This fills the gap between the power-constrained environment of the bracelet and the cloud-deployed software server. The clinician associated with the patient can consult her reading history through a web interface. Clinicians can sign up to receive real-time alerts in case the system detects the onset of hypertension. The present section presents the hardware and software components and details the design decisions behind the system architecture shown in Figure \ref{fig:architecture}.

\subsection{Hardware Components}
The wearable bracelet device is comprised of innovative biosensors that monitor the user's blood pressure. They measure the microfluidic channel deformation based on the impedance principle. Their large size covers a sufficient area of the wrist to detect the small changes in pressure as blood pushes against the wall of the radial artery. The sensor is connected to an electronic module comprised of a lock in amplifier, a microcontroller, a current generator, a screen, battery and a USB port. 

The current generator is connected to the electrodes of the sensor used to measure the impedance of the deformable microfluidic channel. The lock in amplifier uses the sensor voltage signals from which noise is extracted. The signal is converted from analog to digital, after which it is forwarded to the microcontroller. The microcontroller is responsible for sensor data acquisition, data encryption and wireless transmission to the associated user device. The microcontroller encrypts the waveform information and transmits it to the paired device over Bluetooth. The bracelet includes a screen used to pair it with a mobile device. The power supply for the integrated electronic modules is provided through the use of a battery, which can be recharged using a USB port.

\subsection{Software Components}
The software components are spread across the bracelet, a mobile application with Android and iOS implementations and the software server. The bracelet includes the required software to record and carry out initial analysis on the pressure waveform \cite{17}, be paired with a mobile device and send encrypted reading data to it. The mobile application acts as the bridge between the bracelet and the software server. Its inclusion reduces the bracelet's weight, complexity and cost, as it only requires a BLE chip for communication. Furthermore, it keeps end users in control of the recorded data, which can be viewed and managed through the mobile application. 


The bracelet's sensing layer is comprised of several dedicated sensors that use the arterial tonometry technique \cite{kemmotsu1991arterial} to calculate hemodynamic parameters. When equipped, the dedicated sensors are positioned along a peripheral artery (usually the radial artery) which they make contact with without flattening it or disturbing blood flow. The waveform of the pulse is captured by the sensor and sent to the microcontroller, which discards corrupt data sequences caused by incorrect placement of the bracelet. Correct waveforms are sent to the paired mobile device. The accompanying application determines minimum and maximum values based on the Hill Climbing method \cite{rawat2013}. The algorithm is implemented at mobile device level to maximize the bracelet's battery life. The most important parameters obtained are the systolic and diastolic blood pressure values. 

\begin{figure}[t]
    \centering
    \includegraphics[width=\linewidth]{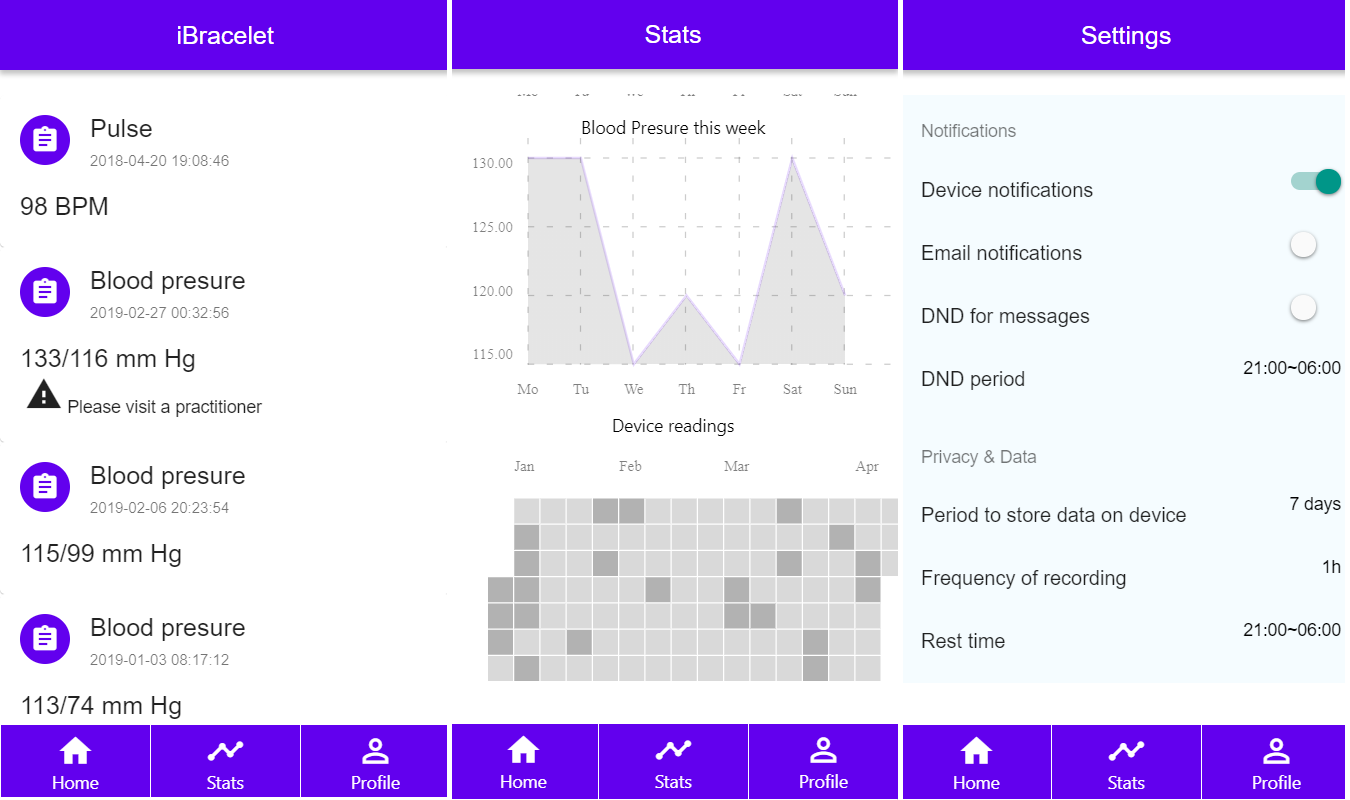}
    \caption{Mobile application user interface screens}
    \label{fig:appui}
\end{figure}

Users can monitor blood pressure values using the mobile application or through the software server's web interface. Both offer the possibility of registering family members or physicians, giving them access to data. Physicians have the possibility of receiving real-time alerts at the onset of persistent hypertension. As shown in Figure \ref{fig:appui}, the application also records heart rate. The user can configure the measurement frequency, mark the period of time they reserve for resting and configure how they choose to be notified. The settings are designed to provide balance between the frequency of measurement, device battery life and avoiding the psychological observer effect. Furthermore, values recorded during the resting period are important, as they might signal increased risk \cite{oded09}.

The risk of preeclampsia is present when persistent hypertension is detected. This is defined as systolic values over 140 mmHg or diastolic values over 90 mmHg recorded over two measurements at least 6 hours apart \cite{english2015risk}. In this case, the mobile application can raise an alert to notify the end user to contact their physician. All recorded values are stored locally by the mobile application. Once an Internet connection is available, they are sent to the application server.

The server is the central repository where all recorded blood pressure values are stored. End-users and physicians access the server through a web application that enforces role-based access to ensure data privacy. The application allows all users to visualize recorded values, statistical data and alert history. The server also carries out the final, but most complex phase of data processing. Physicians can enter additional contextual and clinical information into the system such as patient age, weight, height, race, smoking status and cholesterol level. These values are fed to a machine learning classifier - one of the project's innovations on the software side. The classifier was trained using the open data set from the Massachusetts University Amherst and National Health and Nutrition Survey and is described in \cite{17}. It provides a finely tuned risk profile for hypertensive conditions in pregnancy and identifies those cases that are most prone to preeclampsia, before physiological changes (e.g. proteinuria) can be detected.



\subsection{Technical Challenges}
The most important challenge regarded the accuracy of the waveform-based measurement in the context of a permanently-equipped bracelet. The research team considered placing the device both on the arm, like a regular blood pressure cuff and on the wrist, similar to a watch. 

The main trade-off concerns the fact that the arm is closer to heart and at the same level, which increases the likelihood of accurate readings; since technical evaluation showed that a wrist-mounted bracelet produces accurate readings, it was preferred. This also has the added advantage of not looking out of place, as the device resembles a fitness tracker\footnote{Further details are subject to patent protection.}. Additional optimizations were carried out to minimize battery consumption and ensure that normal wear and tear does not affect reading accuracy or result in leaks from the microfluidic layer. Dividing software components across the bracelet, mobile device and server complements hardware design choices, lowers entry cost and allows additional features related to disease identification and early warning to be evaluated and rolled out on the server side. 

\section{Conclusions}
\label{sec:conclusion}
Hypertensive disorders can appear at any time during pregnancy, while preeclampsia is usually reported after 20 weeks. Early detection is paramount to give the mother the opportunity to receive proper treatment and minimise risks during pregnancy and delivery. The platform described is part of the \textit{i-bracelet} project \cite{16}, which aims to contribute towards decreasing hypertension related risks in pregnant women. The current paper is focused on the software components, while hardware aspects are discussed in our previous work \cite{17,marin2018,marin19}.

When compared with existing approaches, our proposal provides an end-to-end solution for continuous, real-time monitoring that combines an innovative, easy to use piece of hardware integrated with software services that enable data collection, advanced data analysis to predict future cases of persistent hypertension together with statistical reporting.

From the hardware side, we believe the greatest improvement will be to extend the project's scope in order to cover prevention and management of hypertension in the general population. The best way to achieve this is to integrate the bracelet's technology into well-known wearable devices. In the interim, we aim to integrate recordings with globally used platforms such as Apple HealthKit\footnote{https://developer.apple.com/healthkit/} and Google Fit\footnote{https://www.google.com/fit/}. On the server side, we have evaluated several machine learning algorithms and obtained promising results \cite{17,marin19}. We intend to further improve classification and increase its predictive power, both in the particular case of pregnant women and within the general population.

\section*{Acknowledgement}
\noindent This work was funded by a grant of the Romanian National Authority for Scientific Research and Innovation, CCCDI-UEFISCDI, project number 59/2017, Eurostars Project E10871, i-bracelet-\textit{"Intelligent bracelet for blood pressure monitoring and detection of preeclampsia"}.

%
%
%
\bibliographystyle{splncs04}
\bibliography{references}
\end{document}